\documentclass[useAMS,usenatbib]{mn2e}
\usepackage{graphicx}
\usepackage{natbib}
\usepackage{longtable}
\usepackage{times}

\title[The host of the SLSN PTF12dam]{A young stellar environment for the super-luminous supernova PTF12dam\thanks{Based on data taken under programs GTC67-13B, GTC69-14A (PI C. Th\"one) and GTC48-14A (PI A. de Ugarte Postigo)}}
\author[C. C. Th\"one et al.]{C. C. Th\"one$^{1}$ \thanks{E-mail:
cthoene@iaa.es}, A. de Ugarte Postigo$^{1,2}$, R. Garc\'ia-Benito$^{1}$, G. Leloudas$^{3,2}$, S. Schulze$^{4,5}$, \newauthor R. Amor\'in$^{6}$ \\
$^{1}$ Instituto de Astrof\'isica de Andaluc\'ia, Glorieta de la Astronom\'ia s/n, 18008 Granada, Spain \\
$^{2}$ Dark Cosmology Centre, Niels-Bohr-Institute, University of Copenhagen, Juliane Maries Vej 30, 2100 K\o benhavn \O, Denmark\\
$^{3}$ Department of Particle Physics \& Astrophysics, Weizmann Institute of Science, Rehovot 76100, Israel\\
$^{4}$ Instituto de Astrof\'isica, Facultad de F\'isica, Pontificia Universidad Catolica de Chile, 306, Santiago, Chile\\
$^{5}$ Millennium Institute of Astrophysics, Vicu\~{n}a Mackenna 4860, 7820436 Macul, Santiago, Chile\\
$^{6}$ INAF - Osservatorio Astronomico di Roma, via Frascati 33, 00040 Monteporzio Catone, Roma, Italy
}

\begin{document}

\date{Accepted. Received; in original form }

\pagerange{\pageref{firstpage}--\pageref{lastpage}} \pubyear{2014}

\maketitle

\label{firstpage}

\begin{abstract}
{The progenitors of super luminous supernovae (SLSNe) are still a mystery. Hydrogen-poor SLSN hosts are often highly star-forming dwarf galaxies and the majority belongs to the class of extreme emission line galaxies hosting young and highly star-forming stellar populations. Here we present a resolved long-slit study of the host of the hydrogen-poor SLSN PTF12dam probing the kpc environment of the SN site to determine the age of the progenitor. The galaxy is a ``tadpole'' with uniform properties and the SN occurred in a star-forming region in the head of the tadpole. The galaxy experienced a recent star-burst superimposed on an underlying old stellar population. We measure a very young stellar population at the SN site with an age of $\sim$\,3\,Myr and a metallicity of 12+log(O/H)$=$8.0 at the SN site but do not observe any WR features. The progenitor of PTF12dam must have been a  massive star of at least 60 M$_\odot$ and one of the first stars exploding as a SN in this extremely young starburst.}

\end{abstract}

\begin{keywords}
supernovae: individual: PTF12dam; galaxies: abundances, starburst
\end{keywords}

\section{Introduction}

Superluminous supernovae (SLSNe) are a recently discovered class of stellar explosion with peak magnitudes of $<$\,--21\,mag \citep[for a review see][]{Gal-YamSci}. Their hosts are predominantly low-luminosity blue compact dwarfs (BCDs) first monitored by non-targeted wide-field surveys, so they went undetected until 2005 \citep[SN\,2005ap;][]{Quimby07}, although some events were later reclassified as SLSNe \citep{Gal-YamSci}. 

SLSNe are divided into two classes: SLSNe Type II have hydrogen in their spectra and they often resemble Type IIn SNe. Their energy probably stems from interactions with the circumstellar medium or previously ejected H-rich material \citep{Gal-YamSci}. Type I are hydrogen-poor with late spectra similar to broad-lined Type Ic SNe \citep{Pastorello10, QuimbyNature}. Their energy is emitted at large radii suggesting interactions with pre-explosion shells \citep{Chevalier11}  \citep[similar to what has been suggested for a gamma-ray burst in 2010;][]{ThoeneNature}. Alternatively, they could be powered by a magnetar \citep[e.g.][]{Kasen10} or they are pair-instability SNe \citep{Gal-Yam09}.

The hosts of SLSNe are distinctively different than other SN hosts. Only recently \cite{Lunnanhosts} and \cite{LeloudasSUSHIES} published two studies of SLSN host samples. They conclude that Type I SLSN hosts are consistently low-luminosity low-mass objects with low metallicities and high specific star-formation rates (SSFRs) while Type II hosts are more massive and metal-rich \citep{LeloudasSUSHIES}, pointing to a different progenitor. While \cite{Lunnanhosts} claim a similarity between long GRB (LGRB) and SLSN-I hosts, \cite{LeloudasSUSHIES} argue that SLSN-I hosts are often even more extreme and belong to the class of extreme emission line galaxies  \citep[EELGs, see e.g.][]{AmorinEELGs}.

In this paper we present a detailed analysis of different parts of the host of PTF12dam and derive conclusions on the progenitor star. PTF12dam at z$=$0.107 was detected by the Palomar Transient Factory on April 20 2012 \citep{QuimbyATEL} and shown to be a hydrogen-poor SLSN similar to SN\,2007bi \citep{Nicholl13}. Its host is a ``tadpole galaxy'' \citep{SanchezAlmeida} with a compact core consisting of several SF regions, where the SN was located, and a fainter tail. Global spectra of the host have been presented by \cite{ChenPTF12dam, Lunnanhosts} and \cite{LeloudasSUSHIES} showing very strong emission lines and a low metallicity which point to recent starburst. Throughout the paper we use a Planck Cosmology with $\Omega_m=$0.315, $\Omega_\Lambda=$0.685 and H$_0=$67.3. 

\vspace{-0.2cm}
\section{Observations}\label{sect:obs}
\begin{figure}
   \centering
   \includegraphics[width=\hsize]{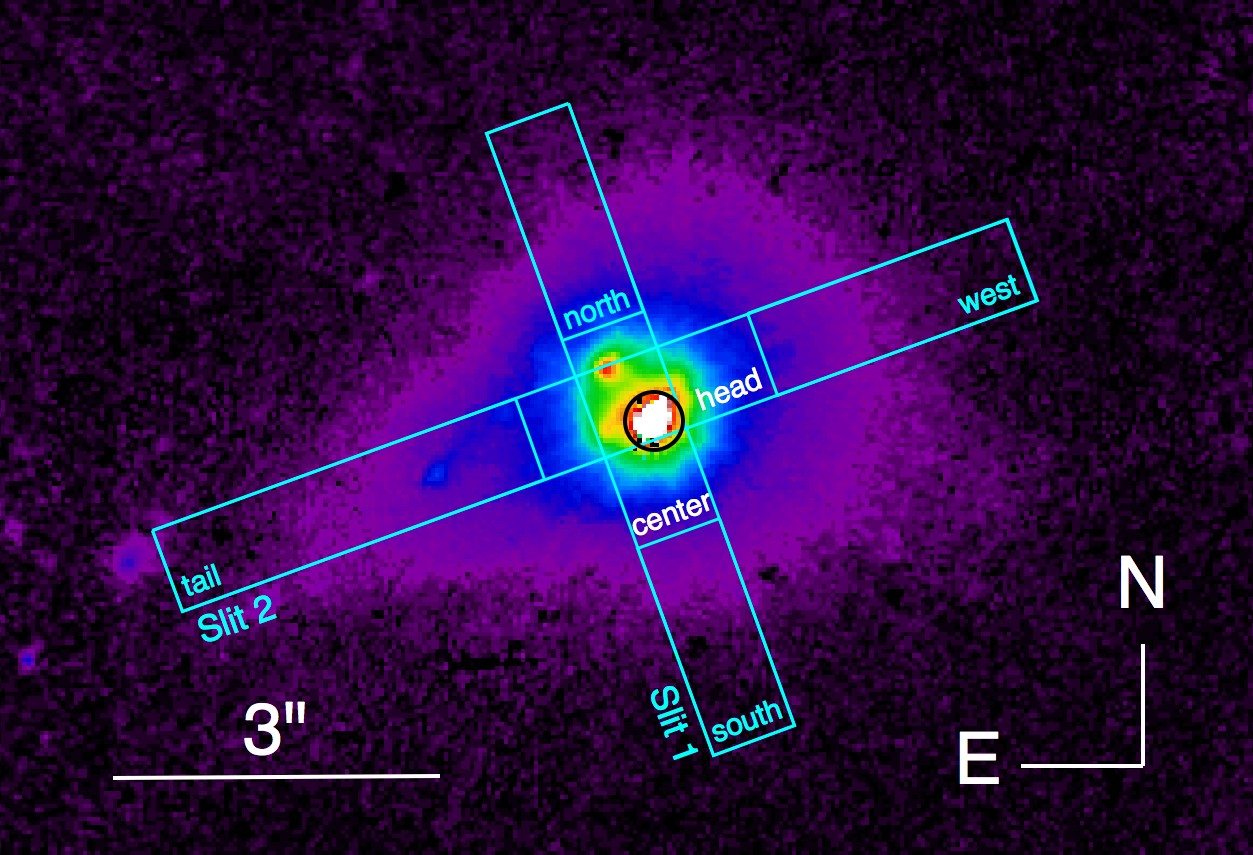}
      \caption{Image of the galaxy with the two slit-position and the extracted regions, the SN position is marked with a black circle. The image was taken from the HST archive (program 12524, PI R. Quimby) while the SN was still present, so the SF regions around the SN site cannot be distinguished.}
         \label{fig:FC}
   \end{figure}

We obtained longslit spectra using OSIRIS at the GTC \citep{Cepa} at two different slit angles (see Fig. \ref{fig:FC}) on Feb. 28 2014 and Apr. 30 2014.  Slit1 was put at a parallactic angle covering the SN position while slit 2 was placed across the tadpole and also covers the SN site. At each epoch, we used the R2000B and R2500R grisms with a combined wavelength range from 3950 to 7700 \AA{} and which provide resolutions of 2165 (0.85\,\AA{}/pix) and 2475 (1.04\,\AA{}/pix) respectively. Exposure times were  3$\times$400\,s at the first epoch and 4$\times$400\,s at the second epoch. Relative flux calibration was obtained by observing the standard stars G191 and GD143 at epoch 1 and 2.  The seeing was good with 1.2 and 0.9 arc sec in the two nights respectively. The spectra are then divided into 3 spatial bins each and analyzed separately. The bins have a width of 8--12 pixels to cover different parts of the galaxy (see Fig.\ref{fig:FC}). We also extract the integrated spectrum for both slit positions. 

On August 4, 2014, we performed tunable narrow-band filter observations with OSIRIS/GTC of the field around PTF12dam. Observations consisted of $5\times750$\,s exposures using a 12\,\AA{} filter in steps of 8\,\AA{} around the central wavelength of H$\alpha$ at the redshift of PTF12dam. An additional 200\,s continuum image was taken, scaled to the flux of each narrow band filter and subtracted from the frames after PSF matching.

\vspace{-0.5cm}
\section{Results}
\begin{figure*}
   \centering
   \includegraphics[height=16.2cm,angle=90]{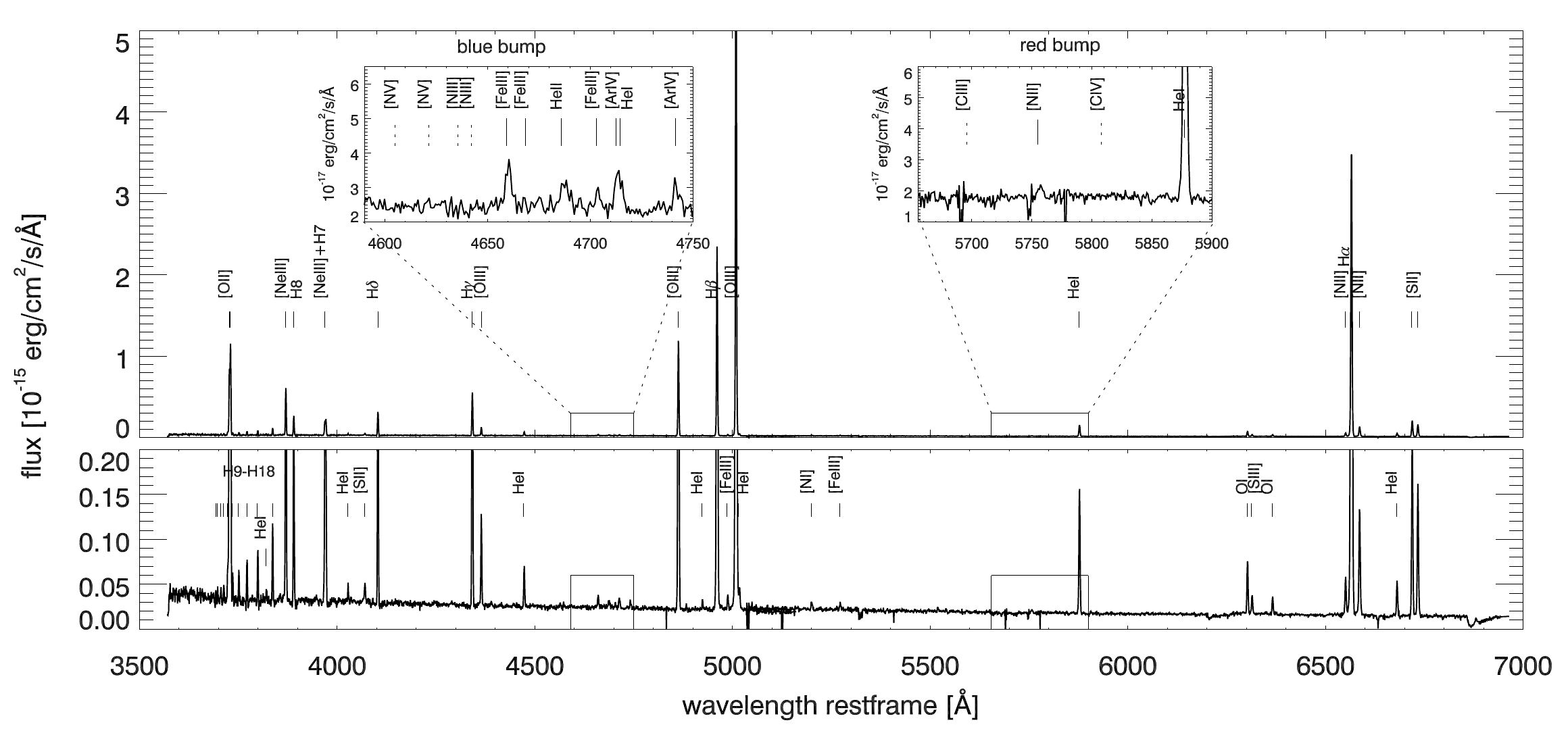}
      \caption{Entire spectrum at the SN position (center position of slit 1). The two insets show the nebular lines (solid lines) and non-detected WR lines (dashed lines) in the region of the blue and red WR bump.}
         \label{fig:WR}
   \end{figure*}

Both spectra show a multitude of emission lines (see Table \ref{table:emission} and Fig. \ref{fig:WR}). In addition to the strong nebular and auroral lines of [O\,II], [O\,III], H$\alpha$, [N\,II] and [S\,II], we detect 16 lines of the Balmer series in emission, several transitions from He II, [Ne\,III], [Fe\,III] and [Ar\,IV], the T$_e$ sensitive [O\,III] line $\lambda$ 4363, [O\,I] $\lambda$ 6300 and most of the permitted transitions of He I. [Ne\,III]/[O\,III] is constant with oxygen abundance due to their similar ionization structure, but [Ne\,III] is often missed because of its lower line strength. He II, [Ar\,IV] and [Fe\,III] require a hard radiation field to be ionized as it is usually provided by WR stars, but other sources are possible. [O\,I] is indicative of shocks and can be used to distinguish between excitations by AGNs and HII regions. The location in the diagnostic diagram of log(OI/H$\alpha$) vs. log([O\,III]/H$\beta$) confirm its origin in a normal HII region \citep[see e.g.][]{Kewley01}.

The host is an EELG as are $\sim$50\% of SLSN-Type I hosts \citep{LeloudasSUSHIES}. With EWs of [OIII]$\lambda$5008 $=$~--945$\pm$2\,\AA{} and H$\alpha$$=$\,--814$\pm$9\,\AA{} (rest-frame) in the integrated spectrum of the galaxy, it is the most extreme example in the sample of \cite{LeloudasSUSHIES}.

\begin{table*}
\scriptsize
\caption{Properties along the two slit positions. N2 and O3N2 metallicities are from \citealt{Marino13}, the ionization parameter U is derived from the models described in \citealt{PM14} using several strong emission lines. For the extinction from the Balmer decrement we use a linear fit to the ratio from H$\alpha$, H$\beta$, H$\gamma$ and H$\delta$ and assume a case B recombination and temperatures and densities of 12000K / 100 cm$^{-3}$. EWs are in rest frame.}          
\label{table:prop}     
\centering                       
\begin{tabular}{l l l l l l l l l}        
\hline               
					&slit1  			&  			&  				&  				&slit2 			& 				&  				&  		\\
					&integrated		&center 		&north 			&south 			& integrated		& head			& tail				&west 	\\  \hline
12 + log(O/H) (T$_e$)  & 8.06 $\pm$ 0.01   & 8.06 $\pm$ 0.01   & 8.15 $\pm$ 0.07   & 8.12 $\pm$ 0.14   & 8.08 $\pm$ 0.03   & 8.05 $\pm$ 0.03   & 7.86 $\pm$ 0.13                 & ---\\
12 + log(O/H) (O3N2)   & 8.07 $\pm$ 0.18   & 8.05 $\pm$ 0.18   & 8.06 $\pm$ 0.18   & 8.07 $\pm$ 0.18   & 8.03 $\pm$ 0.18   & 8.03 $\pm$ 0.18   & 8.05 $\pm$ 0.18   & 8.06 $\pm$ 0.18\\
12 + log(O/H) (N2)    & 8.09 $\pm$ 0.16   & 8.06 $\pm$ 0.16   & 8.07 $\pm$ 0.16   & 8.10 $\pm$ 0.16   & 8.01 $\pm$ 0.16   & 8.02 $\pm$ 0.16   & 8.03 $\pm$ 0.16   & 8.05 $\pm$ 0.16\\
log(N/O) (T$_e$)  & -1.28 $\pm$ 0.07  & -1.32 $\pm$ 0.07  & -1.44 $\pm$ 0.13  & -1.38 $\pm$ 0.30  & -1.39 $\pm$ 0.10  & -1.37 $\pm$ 0.08  & -1.43 $\pm$ 0.37                 &  ---\\
n$_e$[cm$^{-3}$]	([S\,II])     & 109:        & 93:      & 233:      & 212:        & 93 $\pm$ 36       & 95 $\pm$ 31     & 155:                & ---\\
T$_e$[10$^4$K]	([O\,III])   & 1.32 $\pm$ 0.01   & 1.32 $\pm$ 0.01   & 1.28 $\pm$ 0.09   & 1.33 $\pm$ 0.18   & 1.29 $\pm$ 0.03   & 1.32 $\pm$ 0.03   & 1.58 $\pm$ 0.23                 & ---\\
EW H$\alpha$	[\AA{}]	&--814$\pm$9	&--849$\pm$13&--570$\pm$10	&--629$\pm$31	&--605$\pm$9	&--764$\pm$10	&--152$\pm$5		&--133$\pm$3	\\
EW H$\beta$	[\AA{}]	&--146$\pm$4		&--154$\pm$4	&--115$\pm$9	&--105$\pm$14	&--109$\pm$3		&--143$\pm$3		&--34$\pm$1		&--35$\pm$1	\\
E(B-V)   & 0.12 $\pm$ 0.01   & 0.15 $\pm$ 0.01   & 0.03 $\pm$ 0.01   & 0   & 0.17 $\pm$ 0.01   & 0.24 $\pm$ 0.01   & 0.05 $\pm$ 0.03   & 0\\
SFR [M$_\odot$/yr]	H$\alpha$,[OII]&4.3, 4.3	&4.1, 4.2		&0.33, 0.36		&0.13, 0.16		&4.2, 3.6			&5.6, 5.4			&0.15, 0.17		&0.05, 0.07	\\
log(U) & -2.37$\pm$0.14  & -2.37$\pm$0.14  & -2.47$\pm$0.11  & -2.37$\pm$0.11  & -2.32$\pm$0.12  & -2.36$\pm$0.13  & -2.33$\pm$0.14  & -2.45$\pm$0.24\\
 $[$O\,III]/[O\,II]			&5.43$\pm$0.03	&5.70$\pm$0.02&4.50$\pm$0.05	&4.45$\pm$0.11	&6.56$\pm$0.05	&6.71$\pm$0.03	&4.61$\pm$0.11	&3.75$\pm$0.28\\ \hline
\end{tabular}
\end{table*}

\vspace{-0.5cm}
\subsection{Abundances, star-formation and extinction}
We determine metallicities in the different parts of the galaxy using the strong line parameters of N2 and O3N2 with the most recent calibration of \cite{Marino13} as well as by direct abundance measurements using the electron temperature (T$_e$) sensitive line of [O\,III]$\lambda$4363 (see Table 1). We also determine the relative abundances of N, Ne, Ar, Fe and He (see Table 2 in the Appendix).

The metallicity is generally low with 12+log(O/H)$=$8.04--8.09 or Z$=$$\sim$1/4 Z$_\odot$ and the values from different methods match surprisingly well. The host does not show any enhancement in the N/O ratio as some green pea galaxies (GPs, a subsample of EELGs) \citep{Amorin}, but has abundances of Ne, He, Ar and Fe very similar to those galaxies \citep{AmorinGPabundances}. The ionization level is high and comparable to the most extreme EELGs \citep{JaskotOey}, indicating a hard radiation field and young stellar population \citep[e.g.][]{MartinManjon}. There is very little variation of metallicity and abundances along the slits except for some lower values in the tail of the tadpole, though the values are within the errors of the methods. The extinction is derived using the linear fit to the Balmer decrements of H$\alpha$, H$\beta$, H$\gamma$ and H$\delta$. Some low  extinction is present in the head of the tadpole, where most of the SF takes place, in the rest of the galaxy the extinction is consistent with zero. The Galactic extinction in the line-of-sight is E(B--V)=0.01mag.

The SFR derived from H$\alpha$ \citep{Kennicutt} gives values of 4.3 M$_\odot$/yr for the two slit positions. Most of the SF comes from the tadpole head where the SN exploded. We take the values for the SFR, the magnitude in the B-band and the stellar mass from \cite{LeloudasSUSHIES} that used the spectrum of slit position 1 scaled to the photometry of the entire galaxy. With this we get values of 24\,M$_\odot$/yr/L/L* and 1.39 Gyr$^{-1}$ for the luminosity and mass weighted specific SFR. These are some of the highest values of the sample in \cite{LeloudasSUSHIES} and on the upper end of the distribution for EELGs.  

\vspace{-0.4cm}
\subsection{Age of the stellar population}
Several diagnostics can be used to constrain the age of the underlying stellar population and with that the age and mass of the progenitor star. This is the first time we are able to put very tight constraints on the age of a SLSN progenitor. In the following we refer to the spectra of the galaxy head that cover the SN site itself as we are interested in the age of the stellar population around the SN.

The EW of the H$\alpha$ and H$\beta$ emission lines show a strong dependence on age in the first 10--20\,Myr both for an instantaneous burst (ib) and continuous star-formation (csf), somewhat dependent on the metallicity of the gas. Due to the young age no considerable absorption in the Balmer lines had to be taken into account. At a metallicity of 12+log(O/H)$\sim$8.0 (Z$=$0.004) adopting new models generated by \cite{Levesque12} using Starburst99 \citep{Leitherer99} we get ages of 4 (ib) and 10-15\,Myr (csf).

Another strong age indicator is He\,I \citep{GonzalezDelgado99} which is only present in emission up to 5 Myr after the starburst (ib). The EWs of the He\,I~$\lambda\lambda$3810, 4026, 4471 and 4922 lines consistently give an age of 3\,Myr (ib) or 5--10\,Myr (csf) \citep{GonzalezDelgado99}.

Very young starbursts are can further be recognized by the Balmer continuum showing in emission instead of a Balmer break in absorption \citep{SanchezAlmeidagalaxies}. Higher order Balmer lines are quickly affected by stellar absorption and disappear in emission, e.g. H8 would be dominated by absorption for an age of $>$ 5\,Myr (ib) \citep{GonzalezDelgado99}. We note some low-level absorption in the high order Balmer lines, but the center part of the galaxy clearly shows the Balmer series in emission down to the Balmer break putting a maximum age of 5\,Myr. The SED has a large upturn bluewards of the Balmer break and is detected by {\it GALEX} \citep{ChenPTF12dam}, supporting a very young population.

\begin{figure}
\centering
   \includegraphics[width=\columnwidth]{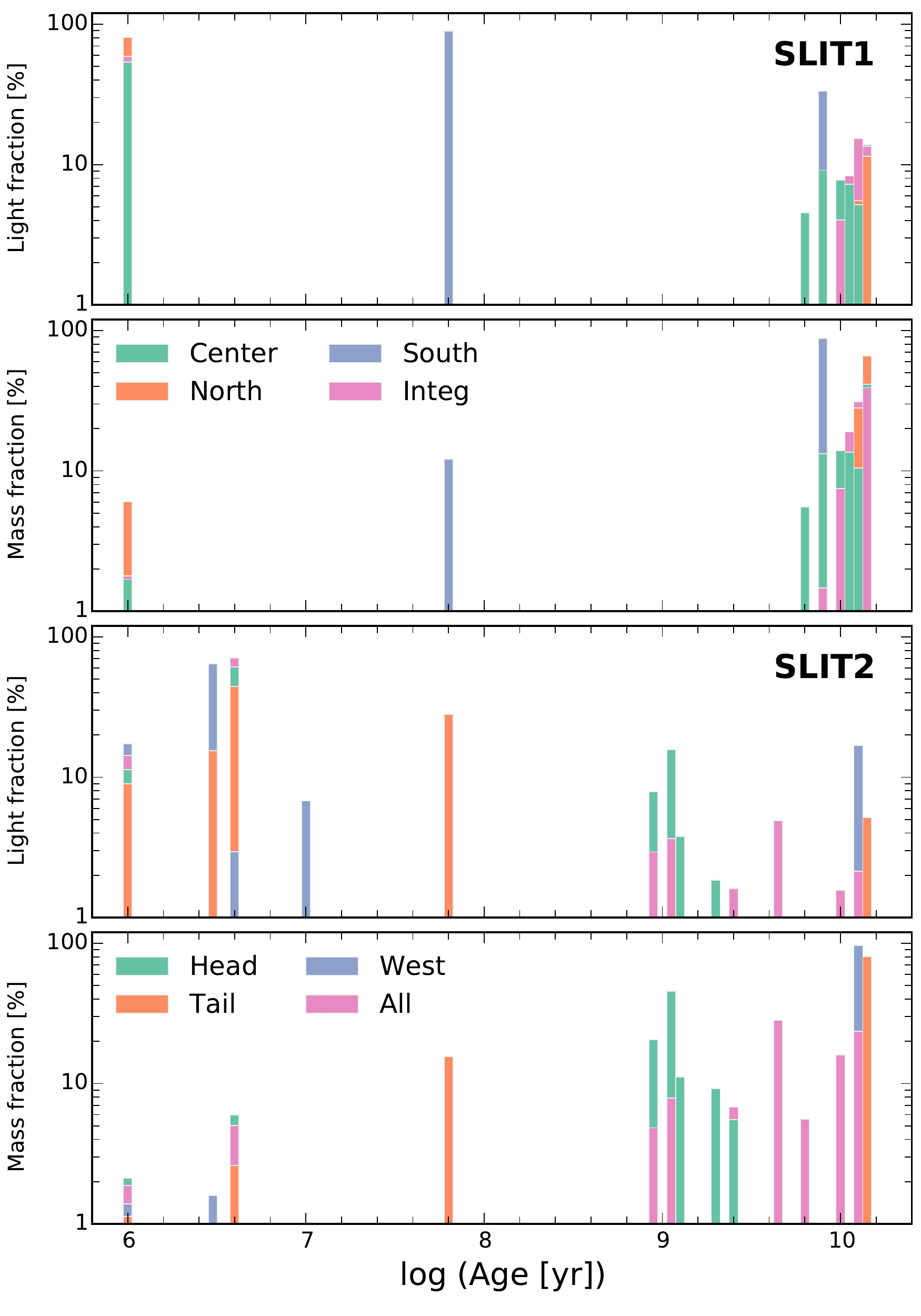}
           \caption{SFH modeling for the two slit positions showing the contributions to mass and luminosity of the different stellar populations. The individual regions are color coded.}
         \label{Fig:SFH}
   \end{figure}

To further investigate the age of the stellar population at the SN site, we perform star-formation history (SFH) modeling using the spectral continuum. We use the {\sc starlight} code \citep{CidFernandes05}, which fits a spectrum with a non-parametric linear combination of single stellar population (SSP) models. Dust effects are modeled using a Calzetti reddening law with R$_V$ = 3.1 and emission lines have been masked out. The SSP base has been built from the Granada \citep{GonzalezDelgado05} and MILES \citep{Vazdekis10} libraries, covering a wide metallicity range (log Z/Z$_{\odot}$ from -2.3 to -0.4), and ages from 0.001 to 14 Gyr\footnote{Details on these libraries can be found in \cite{GonzalezDelgado14}.}. We assume a Salpeter IMF.  All spectra present a very similar SFH (see Fig. \ref{Fig:SFH}): a young component with less that 10 Myr and old component $>\sim$ 1\,Gyr. This has also been found in similar objects with recent star formation like BCDs \citep{PerezMontero10} and GPs \citep{AmorinGPabundances}. An intermediate component is seen in the southern part of slit 1 and the tail spectrum of slit 2.

\vspace{-0.3cm}
\subsection{The larger environment of the PTF12dam host}
The origin for the very recent burst in SF in the host of PTF12dam is not known. To explore the possibility tha SF was triggered by interaction we observed the field around PTF12dam with narrow-band tunable filters centered at H$\alpha$ at $z=$0.107 (see Sect.\ref{sect:obs}). The use of etalon filters results in a shift in wavelength with distance from the optical axis such that with the 5 steps covered in our observations we can detect H$\alpha$ in emission in a field-of-view (FoV) of $\sim$ 2$^{\prime}$ (244\,kpc) around the optical axis.

Within this FoV we detect only 3 emitting sources (see Fig. 1 in the Appendix): G1 is at a distance of 139 kpc with an emission peak at --315 km\,s$^{-1}$ and has an SDSS photo-z of $0.188\pm0.1341$. G2 was also covered by slit 1 but has an emission-line redshift of $z=$0.443. G3 has two emission peaks which could correspond to H$\alpha$ and [N\,II] in a high metallicity object and also has a photo-z of 0.118$\pm$0.064. None of the objects are clearly interacting with PTF12dam, requiring an alternative solution for triggering the star formation episode that we are observing.

\vspace{-0.3cm}
\subsection{Wolf-Rayet features}
We also searched the different spectra for signatures of WR stars (see Fig.\ref{fig:WR}). The spectra have good S/N around the so-called ``blue bump''. The region of the ``red bump'' is affected by sky lines which excludes the detection of [C\,IV]. The WR features in the blue bump consist of broad lines of [N\,V] $\lambda\lambda$ 4605,4621 and [N\,III] $\lambda\lambda$ 4641, 4635, while [C\,III] and [C\,IV] make up the red bump. In addition, narrow nebular lines of  [Fe\,III] $\lambda\lambda$ 4659, 4668, 4702, [Ar\,IV] $\lambda\lambda$ 4712, 4742 and He II $\lambda$ 4686 can be detected in the blue and He I $\lambda$ 5876 in the red bump. In the presence of WR features, He II should show both a narrow nebular and a broad WR component. 

We detect all of the narrow nebular lines in both the blue and the red bump in the integrated spectra of both slit positions, but there seems to be neither the broad WR lines nor a broad component in He II line present (see Fig. \ref{fig:WR}). All of the flux in these nebular lines comes from the bright SF regions in the tadpole head, the S/N of the other regions are not high enough to detect any of the lines except He I in the red bump. 

From the non-detection of WR lines we derive limits on the numbers of WR stars following the prescription of \cite{LopezWR} using the spectra of the tadpole head in slitposition 2. The flux limits depend on the width of the WR lines for which we use 1000 and 3000 km\,s$^{-1}$ FWHM as empirical lower and upper limits \cite[see][their Fig. 3]{BKD08}. We then derive luminosities of 2.37--7.16$\times$10$^{39}$erg\,s$^{-1}$ (1000--3000 km\,s$^{-1}$) for the blue and 1.66 -- 4.99$\times$10$^{39}$erg\,s$^{-1}$ for the red bump and limits of $<$\,4175--6475 WNL and $<$\,988 -- 2970 WCE stars.

\vspace{-0.3cm}
\subsection{Kinematics}
Using intermediate resolution spectra \cite{Amorin12} found that the emission lines of GPs consist of a narrow and a broad component with a FWHM of up to 600 km\,s$^{-1}$. Such components have also been detected in other compact starbursts \citep[see e.g.][]{Izotov07, James} and extragalactic HII regions \citep[e.g.][]{Firpo11} and might trace outflows from WR/stellar winds or SN explosions. We analyze the structure of H$\alpha$ and [O\,III] using the {\sc ngaussfit} routine in IRAF (see Fig. \ref{fig:kinematics} in the Appendix). 

The emission lines show surprisingly little kinematic features. The emission lines across the tadpole show no sign of rotation and the centroids differ only by 28 km\,s$^{-1}$ across the slit, although this might be an issue of spectral resolution. All lines are best fit with a combination between a narrower and a wider component, comparison with a single Gaussian fit results in larger residuals that are positive in the wings indicating that the broad component is actually present. The narrow components are barely resolved with widths of 120\,--\,140 km\,s$^{-1}$ FWHM (our spectra have resolutions of $\sim$ 100 km\,s$^{-1}$) while the broad components have a FWHM of 180 -- 200 km\,s$^{-1}$. The two components do not show a large shift in velocity except probably for [O\,III] in slit 2 while \cite{Amorin12} found considerable shifts compared to the line centroid for GPs. Full width zero intensity (FWZI) of the emission lines are $\sim$ 400 km\,s$^{-1}$ which is larger than the rotation in a dwarf galaxy and might be an indication for stellar winds or outflows.

\vspace{-0.5cm}
\section{Discussion}
The host of PTF12dam is a very young starburst even exceeding most EELGs and GPs in terms of [O\,III] EW, SSFR and ionization. It is also one of the most star-forming SLSN hosts detected so far and the only one showing He I emission, although this might be an effect of S/N. However, it does not have an extremely low metallicity and SP modeling clearly shows that this galaxy did not get formed recently. It has an older population of several Gyr followed by a long, quiet period and a recent SF epoch with an age of a few Myr. The properties are largely uniform throughout the galaxy while \cite{SanchezAlmeida} observe abundances up to 0.5 dex higher in the tail of nearby tadpoles. The tail of the PTF12dam host might, however, have a somewhat older SP than the tadpole head, but the metallicity is similar. Galaxy interaction is largely excluded as trigger for the recent starburst and infact most tadpoles are isolated galaxies. \cite{SanchezAlmeida} propose the inflow of cold, metal-poor gas onto the tadpole head as likely SF trigger which could be an appealing option for the PTF12dam host.

It is curious that, despite its young age, we do not detect WR features in the host of PTF12dam while the spectrum shows strong He II lines. However, the fraction of starbursts with WR features drops with metallicity and at 12+log(O/H)$<$ 8.0 only 50\% show WR features \citep{Shirazi12}. This might be due to lower wind strength at low metallicity or a more homogeneous evolution of massive stars. Spatially resolved studies of WR galaxies infact show that regions with He II emission and those with WR features can be spatially separated \citep{KehrigWR}, hence WR stars might not be the only cause for producing He II emission. WR features have not been detected in all GPs \citep{JaskotOey} although this could be an issue of S/N. Our limits on the number of WR stars indicates that the features might simply be too weak to be detected in such a galaxy at that redshift or the starburst could simply be still too young to show strong WR signatures.

\cite{Lunnanhosts} concluded that Type I SLSN hosts belong to a similar population than LGRB hosts although SLSN hosts tend to have even lower masses than LGRB hosts. In contrast, \cite{LeloudasSUSHIES} proposed that hydrogen-poor SLSNe are found on average in younger locations than GRBs. Some nearby LGRB hosts do show evidence for WR features \citep{Han}. Metallicities are usually sub solar but all of them have 12+log(O/H)$>$8.0 and the sites of z$<$0.3 LGRB hosts even have 12+log(O/H)$>$8.2 \citep{Levesque10}. Most Type I SLSN host, in contrast, have metallicities of $\sim$1/4 solar with some even going down to 1/10 solar metallicity or less. 

For the first time we are able to set strong limits on the age of a SLSN progenitor by studying its immediate environment and hence on its mass. SP modeling and high excitation lines suggests that the SLSN was formed in the most recent starburst with an age of $\sim$3\,Myr corresponding to a star of $>$60 M$_\odot$ \citep[e.g.][]{MeynetMaeder}. Based on the H$\beta$ EW of SN sites \cite{Sanders} determined the progenitor mass for different types of stripped-envelope SNe and GRBs concluding that broad-line Ic SNe and GRBs have the youngest and most massive progenitors. Others have found similar ages of 5--6 Myr for the SP surrounding GRB sites using high angular resolution photometry, integral-field spectra and SP modeling \citep{Ostlin, Thoene08, Thoene14, Christensen08}. IFU observations of SN locations show that SN Ibc progenitors are stars with masses of $<$40\,M$_\odot$ with some possible exceptions \citep{Kunca13}.

PTF12dam has a lower the age and higher mass than other SN progenitors suggesting that SLSNe might be the most massive stars and those to explode first after the onset of a star-burst, even before GRBs. \cite{Nicholl13} proposed a magnetar as possible driver for the high luminosity of PTF12dam, but new observations would require special parameters for this model to work \citep{ChenPTF12dam}. A CSM interaction model provides a good fit to the light curve but is considered problematic since it requires a very large ejecta mass and a H-poor CSM \citep[29 and 13 M$_\odot$ respectively][]{ChenPTF12dam}. The mass limits derived here are, however, consistent with this possibility. Whatever the exact model for this SN, its progenitor was likely a single, very massive, star created in a recent star-burst and one of the first stars of that starburst that exploded as a SN. Further high angular resolution studies are needed to determine the general population of hydrogen poor SLSN progenitors.

\vspace{-0.7cm}
\section*{Acknowledgments}
Based on observations made with the GTC, installed in the Spanish ORM of the IAC on the island of La Palma. CCT and AdUP acknowledge support from a Ram\'on y Caj\'al fellowship of the Spanish Ministry of Economy and Competitivity and partial support from AYA2012-39362-C02-02. AdUP further acknowledges support by FP7-PEOPLE-2012-CIG 322307. DARK is funded by the DNRF. RGB acknowledges support from MICINN AYA2010-15081, StS from CONICYT-Chile FONDECYT 3140534, Basal-CATA PFB-06/2007 and IC120009. RA acknowledge the FP7 SPACE project ASTRODEEP (Ref. No: 312725) of the EC.

\vspace{-0.5cm}
\footnotesize{

}

\appendix
\section{Supplementary material}

\newpage
\begin{table*}
\caption{Line fluxes in the individual regions and the combined spectra for each slit. Fluxes are in units of 10$^{-16}$ erg/cm$^2$/s and not corrected for extinction.}             
\label{table:emission}      
\centering                          
\begin{tabular}{l l l l l l l l l}        
\hline                 
line			&	slit1		 & 				 &  &  					&slit2 & &  &  \\
			&	integr.		&head		 & north &south 			&integr.& head&tail&west \\  \hline
H-18			&---			&0.25$\pm$0.09	&---&---					&--- &0.19$\pm$0.13 &--- &---\\
H-17			&---			&0.29$\pm$0.09	&---&---					&--- &0.29$\pm$0.09 &--- &---\\
H-16			&---			&0.41$\pm$0.11	&---&---					&0.34$\pm$0.21 &0.31$\pm$0.11 &--- &--- \\
H-15			&---			&0.52$\pm$0.09	&---&---					&0.39$\pm$0.16 &0.42$\pm$0.14 &---  &--- \\
H-14			&---			&1.16$\pm$0.27	&---&---					&1.05$\pm$0.30 &0.93$\pm$0.26 &--- &--- \\
$[$OII$]$ 3727&28.2$\pm$0.9 	&23.4$\pm$0.44	&3.19$\pm$0.17&1.70$\pm$0.19&19.3$\pm$0.9   &17.4$\pm$0.36   & 1.98$\pm$0.22   & 0.71$\pm$0.23  \\
$[$OII$]$ 3729&37.7$\pm$0.8	& 30.4$\pm$0.39	&4.83$\pm$0.19&2.11$\pm$0.18 &29.5$\pm$1.0 &26.5$\pm$0.38 &2.04$\pm$0.19  & 1.05$\pm$0.24 \\
H-13			&1.25$\pm$0.52&0.77$\pm$0.23	&--- & ---					&0.68$\pm$0.25 &0.54$\pm$0.24 &--- &---\\
H-12			&0.96$\pm$0.23& 0.83$\pm$0.14	&--- & ---					&0.66$\pm$0.15 &0.76$\pm$0.08  &--- &---\\
H-11			&1.39$\pm$0.23&1.29$\pm$0.13 	&--- &--- 					&1.09$\pm$0.21 & 1.00$\pm$0.09 &--- &--- \\
H-10			&1.96$\pm$0.25&1.54$\pm$0.15 	&--- & ---					&1.50$\pm$0.15 &1.57$\pm$0.07  &--- &--- \\
HeI 3819		&---			&0.34$\pm$0.13	 & ---& ---					& ---&0.32$\pm$0.11  &--- &--- \\
H-9			&2.95$\pm$0.20&2.43$\pm$0.09 	&0.44$\pm$0.10& ---		&2.22$\pm$0.21 &1.93$\pm$0.13  &0.23$\pm$0.09 &--- \\
$[$NeIII] 3868	&20.0$\pm$0.28&24.2$\pm$0.10 	&2.27$\pm$0.09&1.05$\pm$0.09&15.8$\pm$0.26 &14.3$\pm$0.21 &1.08$\pm$0.09 &0.46$\pm$0.07\\
H-8			&7.93$\pm$0.28&6.69$\pm$0.12 	&0.89$\pm$0.10&0.28$\pm$0.08&6.28$\pm$0.16 &5.57$\pm$0.11 &0.42$\pm$0.07 & 0.17$\pm$0.07 \\
$[$NeIII] 3968	&6.10$\pm$0.36&5.69$\pm$0.27	&0.57$\pm$0.11&0.18$\pm$0.09&4.24$\pm$0.22 &3.95$\pm$0.15  &0.26$\pm$0.07 &0.13$\pm$0.06  \\
H$\epsilon$	&6.28$\pm$0.34&4.78$\pm$0.22	&0.72$\pm$0.12&0.46$\pm$0.09&5.37$\pm$0.24 & 5.03$\pm$0.17 &0.25$\pm$0.08 &0.22$\pm$0.06  \\
HeI 4026		&0.63$\pm$0.29&0.38$\pm$0.07 	&--- & ---					&0.63$\pm$0.14 &0.54$\pm$0.07  &--- & --- \\
$[$SII] 4068	&0.81$\pm$0.13&0.89$\pm$0.08 	&--- &--- 					&0.48$\pm$0.13 &0.45$\pm$0.06  &--- &--- \\
H$\delta$		&10.6$\pm$0.19&8.87$\pm$0.17 	&1.21$\pm$0.06&0.53$\pm$0.07&8.88$\pm$0.14 &8.19$\pm$0.12  &0.56$\pm$0.06 & 0.25$\pm$0.05 \\
H$\gamma$	&19.7$\pm$0.09&16.3$\pm$0.15 	&2.16$\pm$0.08&1.08$\pm$0.06&16.9$\pm$0.18 & 15.4$\pm$0.09 &1.09$\pm$0.06 &0.37$\pm$0.04 \\
$[$OIII] 4363	&3.64$\pm$0.09 &0.36$\pm$0.01 	&0.36$\pm$0.06&0.19$\pm$0.06&2.97$\pm$0.15  &2.80$\pm$0.10&0.30$\pm$0.09 &--- \\
HeI 4471		&1.59$\pm$0.13&1.47$\pm$0.08 	&--- &--- 					&1.36$\pm$0.08 &1.22$\pm$0.05  &0.15$\pm$0.04 &--- \\
$[$Fe III] 4658 &0.53$\pm$0.09&0.43$\pm$0.07 	&--- &--- 					&0.41$\pm$0.07 &0.36$\pm$0.05  &--- &--- \\
HeII 4686		&0.30$\pm$0.05&0.36$\pm$0.06 	&--- &--- 					&0.34$\pm$0.08 &0.30$\pm$0.05  &--- &---  \\
$[$FeIII] 4702	&0.31$\pm$0.10&0.16$\pm$0.05	&--- & ---					&0.16$\pm$0.07 &0.09$\pm$0.05  & ---& --- \\
HeI 4711\,+\,$[$ArIV] 4713	&0.56$\pm$0.09&0.52$\pm$0.09 	&--- & ---					&0.39$\pm$0.08 &0.43$\pm$0.05  &--- &---  \\
$[$ArIV] 4740	&0.48$\pm$0.10&0.33$\pm$0.09 	&0.10$\pm$0.04 &--- 		&0.19$\pm$0.06 &0.19$\pm$0.06  & ---& ---\\
H$\beta$		&44.9$\pm$0.01&38.1$\pm$0.04 	&4.74$\pm$0.05&2.10$\pm$0.05 &40.1$\pm$0.43 &36.5$\pm$0.33  &2.61$\pm$0.04 &0.92$\pm$0.04 \\
HeI 4922		&0.31$\pm$0.07&0.26$\pm$0.04 	&--- 	&--- 					&0.24$\pm$0.08 &0.27$\pm$0.04  &--- &--- \\
$[$OIII] 4959 	&90.0$\pm$0.09&76.6$\pm$0.09 	&9.21$\pm$0.06&4.25$\pm$0.06&79.3$\pm$0.09 &73.1$\pm$0.10  & 4.54$\pm$0.06& 1.62$\pm$0.04\\
$[$FeIII] 4986	&0.57$\pm$0.11&0.50$\pm$0.06	&---			&			&0.43$\pm$0.12	&0.44$\pm$0.06	&---	&---	\\
$[$OIII] 5008	&267.8$\pm$0.22&230.1$\pm$1.0 	&26.9$\pm$0.16 &12.7$\pm$0.12 &240.8$\pm$3.5 &221.5$\pm$4.4  &14.0$\pm$0.06 &4.98$\pm$0.05  \\
HeI 5015		&2.41$\pm$1.14&5.20$\pm$2.03	&---	&---					&0.86$\pm$0.17	&1.25$\pm$0.46	&0.09$\pm$0.03	&---	\\
$[$NI] 5199	&0.44$\pm$0.11&0.32$\pm$0.07	&--- &--- 					&0.39$\pm$0.13 &0.40$\pm$0.08  &--- &--- \\
$[$FeIII] 5271	&0.28$\pm$0.17&0.25$\pm$0.06 	&--- & ---					&0.20$\pm$0.09 &0.29$\pm$0.05  &--- &---  \\
HeI 5876		&5.36$\pm$0.09&4.58$\pm$0.05 	&0.59$\pm$0.04 &0.21$\pm$0.03 &5.29$\pm$0.07 &5.40$\pm$0.05  &0.29$\pm$0.04 &0.12$\pm$0.04  \\
$[$OI] 6300	&2.21$\pm$0.09&1.91$\pm$0.07 	&0.27$\pm$0.07 & 0.09$\pm$0.04&2.24$\pm$0.07 &2.22$\pm$0.04 &0.17$\pm$0.03 &--- \\
$[$SIII] 6312	&0.72$\pm$0.08&0.69$\pm$0.08 	&0.07$\pm$0.03 & ---		&0.73$\pm$0.06 &0.75$\pm$0.04  &--- &---  \\
$[$OI] 6365	&0.68$\pm$0.11&0.61$\pm$0.05 	& 0.07$\pm$0.02&--- 		&0.58$\pm$0.06 &0.65$\pm$0.04  &--- &---  \\
$[$NII] 6548	&2.32$\pm$0.84&1.03 $\pm$0.47	& 0.14$\pm$0.07&--- 		& 1.10$\pm$0.18&1.16$\pm$0.07  &--- &---  \\
H$\alpha$	&141.5$\pm$0.36&123.2$\pm$0.51 &13.6$\pm$0.11&5.56$\pm$0.13 &130.3$\pm$0.49 &133.5$\pm$0.5  &6.53$\pm$0.08 &1.87$\pm$0.08 \\
$[$NII] 6584		&5.44$\pm$0.55&4.11$\pm$0.53	& 0.47$\pm$0.06& 0.23$\pm$0.09& 3.40$\pm$0.50&3.58$\pm$0.47  &0.19$\pm$0.11 &0.06$\pm$0.03 \\
HeI 6680		&1.53$\pm$0.14&1.29$\pm$0.07 	&--- & ---					&1.43$\pm$0.14 &1.43$\pm$0.06  &--- &---  \\
$[$SII] 6718	&7.82$\pm$0.22&6.64$\pm$0.23 	&0.88$\pm$0.05 & 0.29$\pm$0.04	&7.51$\pm$0.10 &7.57$\pm$0.08  &0.48$\pm$0.04 &0.11$\pm$0.03  \\
$[$SII] 6732	&6.00$\pm$0.32&5.03$\pm$0.15 	&0.74$\pm$0.05 & 0.24$\pm$0.04  &5.69$\pm$0.09 &5.75$\pm$0.08  &0.38$\pm$0.04 &0.05$\pm$0.03\\
\hline                        
\end{tabular}
\end{table*}

\begin{table*}
\caption{
Further properties and abundances in the individual regions and the integrated spectrum not listed in Table 1. n$_e$ and T$_e$ are electron density and temperatures, ICF is the ionization correction factor. Element abundances have been determined using {\sc pyneb} \citep{Pynet}. [O\,II] and [S\,III] temperatures are derived from [O\,III] temperatures using empirical relations. The abundances of O are listed here separately according to ionization state, the total abundance 12+log(O/H) is also listed in Table 1 of the main paper as ``T$_e$ metallicity''. Metallicities determined by empirical parameters (bottom part of the table) are from: R23 \citep{R23}, PP04 \citep{Pettini04} and PM14  \citep{PM14}. R23 has a double solution, the ratios of [N\,II]/H$\alpha$, however, indicate that in all spectra the lower branch is the one to be adopted. 
}
\label{table:emission}      
\centering                          
\begin{tabular}{l l l l l l l l l}        
\hline                 
line&slit1  &  &  &  &slit2 & &  &  \\
	&	integ.&head & north &south &integ.& head& tail&west  \\  \hline
             n([OII]) cm$^{-3}$         & 107$\pm$55         & 137$\pm$34               & ---     & 188:                  & ---                  & ---      & 483:               & ---\\
               T([OII]) 10$^4$K      & 1.23$\pm$0.01      & 1.25$\pm$0.01   & 1.07$\pm$0.04   & 1.11$\pm$0.08      & 1.23$\pm$0.02      & 1.25$\pm$0.02    & 1.29$\pm$0.10               & ---\\
              T([SIII]) 10$^4$K      & 1.31$\pm$0.02      & 1.30$\pm$0.01   & 1.27$\pm$0.09   & 1.32$\pm$0.19      & 1.27$\pm$0.04      & 1.31$\pm$0.03    & 1.58$\pm$0.25               & ---\\
            12+log(O$^+$/H$^+$)      & 7.41$\pm$0.03      & 7.37$\pm$0.02   & 7.67$\pm$0.09   & 7.62$\pm$0.17      & 7.35$\pm$0.05      & 7.35$\pm$0.04    & 7.33$\pm$0.17               & ---\\
         12+log(O$^{2+}$/H$^+$)      & 7.96$\pm$0.02      & 7.96$\pm$0.01   & 7.97$\pm$0.09   & 7.96$\pm$0.19      & 7.99$\pm$0.04      & 7.96$\pm$0.04    & 7.71$\pm$0.18               & ---\\
                    12+log(O/H)      & 8.06$\pm$0.01      & 8.06$\pm$0.01   & 8.15$\pm$0.07   & 8.12$\pm$0.14      & 8.08$\pm$0.03      & 8.05$\pm$0.03    & 7.86$\pm$0.13               & ---\\
            12+log(S$^+$/H$^+$)      & 5.63$\pm$0.03      & 5.60$\pm$0.03   & 5.86$\pm$0.07   & 5.71$\pm$0.13      & 5.63$\pm$0.03      & 5.63$\pm$0.02    & 5.66$\pm$0.11               & ---\\
         12+log(S$^{2+}$/H$^+$)      & 6.07$\pm$0.07      & 6.12$\pm$0.06   & 6.12$\pm$0.32               & ---      & 6.15$\pm$0.08      & 6.14$\pm$0.06                & ---               & ---\\
            ICF(S$^+$+S$^{2+}$)      & 1.36$\pm$0.03      & 1.39$\pm$0.02   & 1.20$\pm$0.07               & ---      & 1.44$\pm$0.06      & 1.41$\pm$0.04                & ---               & ---\\
                    12+log(S/H)      & 6.34$\pm$0.05      & 6.38$\pm$0.05   & 6.39$\pm$0.21               & ---      & 6.42$\pm$0.07      & 6.40$\pm$0.05                & ---               & ---\\
                       log(S/O)     & -1.72$\pm$0.05     & -1.68$\pm$0.05  & -1.75$\pm$0.22               & ---     & -1.65$\pm$0.07     & -1.65$\pm$0.05                & ---               & ---\\
            12+log(N$^+$/H$^+$)      & 6.13$\pm$0.06      & 6.05$\pm$0.06   & 6.22$\pm$0.09   & 6.24$\pm$0.24      & 5.95$\pm$0.08      & 5.97$\pm$0.07    & 5.90$\pm$0.33               & ---\\
                     ICF(N$^+$)      & 4.52$\pm$0.30      & 4.88$\pm$0.22   & 3.03$\pm$0.61   & 3.18$\pm$1.28      & 5.37$\pm$0.62      & 5.08$\pm$0.48    & 3.44$\pm$1.38               & ---\\
                    12+log(N/H)      & 6.79$\pm$0.06      & 6.74$\pm$0.07   & 6.71$\pm$0.13   & 6.75$\pm$0.30      & 6.68$\pm$0.10      & 6.68$\pm$0.08    & 6.44$\pm$0.38               & ---\\
                       log(N/O)     & -1.28$\pm$0.07     & -1.32$\pm$0.07  & -1.44$\pm$0.13  & -1.38$\pm$0.30     & -1.39$\pm$0.10     & -1.37$\pm$0.08   & -1.43$\pm$0.37               & ---\\
        12+log(Ne$^{2+}$/H$^+$)      & 7.30$\pm$0.02      & 7.47$\pm$0.01   & 7.33$\pm$0.12   & 7.28$\pm$0.26      & 7.30$\pm$0.05      & 7.29$\pm$0.04    & 7.00$\pm$0.24               & ---\\
                 ICF(Ne$^{2+}$)      & 1.08$\pm$0.01      & 1.08$\pm$0.01   & 1.10$\pm$0.02   & 1.10$\pm$0.03      & 1.08$\pm$0.01      & 1.08$\pm$0.01    & 1.09$\pm$0.02               & ---\\
                   12+log(Ne/H)      & 7.33$\pm$0.02      & 7.50$\pm$0.01   & 7.37$\pm$0.12   & 7.33$\pm$0.26      & 7.33$\pm$0.05      & 7.32$\pm$0.04    & 7.04$\pm$0.24               & ---\\
                      log(Ne/O)     & -0.73$\pm$0.03     & -0.56$\pm$0.01  & -0.78$\pm$0.14  & -0.80$\pm$0.30     & -0.74$\pm$0.06     & -0.73$\pm$0.05   & -0.82$\pm$0.28               & ---\\
                 12+log(Ar3+/H)      & 5.19$\pm$0.11      & 5.10$\pm$0.12   & 5.51$\pm$0.27               & ---      & 4.87$\pm$0.17      & 4.88$\pm$0.17                & ---               & ---\\
        12+log(Fe$^{2+}$/H$^+$)      & 5.41$\pm$0.13      & 5.40$\pm$0.11               & ---               & ---      & 5.39$\pm$0.12      & 5.34$\pm$0.10                & ---               & ---\\
                 ICF(Fe$^{2+}$)      & 3.68$\pm$0.22      & 3.94$\pm$0.16               & ---               & ---      & 4.30$\pm$0.45      & 4.08$\pm$0.35                & ---               & ---\\
                   12+log(Fe/H)      & 5.97$\pm$0.14      & 6.00$\pm$0.11               & ---               & ---      & 6.02$\pm$0.13      & 5.96$\pm$0.11                & ---               & ---\\
                       log(Fe/O)     & -2.09$\pm$0.14     & -2.07$\pm$0.12               & ---               & ---     & -2.06$\pm$0.13     & -2.10$\pm$0.11                & ---               & ---\\
                   He$^+$/H$^+$    & 0.081$\pm$0.005    & 0.082$\pm$0.005  & 0.090$\pm$0.01  & 0.075$\pm$0.01    & 0.081$\pm$0.011    & 0.084$\pm$0.015  & 0.112$\pm$0.031               & ---\\
                He$^{2+}$/H$^+$    &  6$\times$10$^{-4}$    & 8$\times$10$^{-4}$               & ---               & ---    & 7$\times$10$^{-4}$    & 7$\times$10$^{-4}$                & ---               & ---\\
                           He/H  & 0.081$\pm$0.005  & 0.083$\pm$0.005               & ---               & ---  & 0.082$\pm$0.011  & 0.085$\pm$0.015                & ---               & ---\\
  12+log(O/H) {\tiny R23 lower}      & 8.18$\pm$0.13      & 8.19$\pm$0.13   & 8.17$\pm$0.13   & 8.23$\pm$0.13      & 8.14$\pm$0.13      & 8.16$\pm$0.13    & 8.11$\pm$0.13   & 8.17$\pm$0.13\\
  12+log(O/H) {\tiny PP04 (N2)}      & 8.11$\pm$0.18      & 8.09$\pm$0.18   & 8.10$\pm$0.18   & 8.13$\pm$0.18      & 8.04$\pm$0.18      & 8.05$\pm$0.18    & 8.06$\pm$0.18   & 8.08$\pm$0.18\\
12+log(O/H) {\tiny PP04 (O3N2)}      & 8.03$\pm$0.14      & 8.01$\pm$0.14   & 8.02$\pm$0.14   & 8.04$\pm$0.14      & 7.98$\pm$0.14      & 7.98$\pm$0.14    & 8.01$\pm$0.14   & 8.02$\pm$0.14\\
        12+log(O/H) {\tiny PM14}      & 8.01$\pm$0.06      & 8.01$\pm$0.06   & 8.04$\pm$0.04   & 8.00$\pm$0.06      & 8.05$\pm$0.04      & 8.01$\pm$0.05    & 7.75$\pm$0.09   & 8.07$\pm$0.17\\
                      log(N2O2)     & -1.04$\pm$0.06     & -1.15$\pm$0.06  & -1.14$\pm$0.07  & -1.09$\pm$0.14     & -1.19$\pm$0.05     & -1.18$\pm$0.04   & -1.24$\pm$0.20               & -1.34 $\pm$ 0.20\\
                      log(N2S2)     & -0.40$\pm$0.05     & -0.45$\pm$0.06  & -0.54$\pm$0.06  & -0.36$\pm$0.18     & -0.58$\pm$0.06     & -0.56$\pm$0.06   & -0.65$\pm$0.25  & -0.43$\pm$0.25\\
   log(N/O) {\tiny (N2S2 PM09)}     & -1.37$\pm$0.31     & -1.43$\pm$0.31  & -1.54$\pm$0.31  & -1.32$\pm$0.31     & -1.60$\pm$0.31     & -1.57$\pm$0.31   & -1.68$\pm$0.31  & -1.40$\pm$0.31\\
        log(N/O) {\tiny (PM14)}     & -1.41$\pm$0.14     & -1.44$\pm$0.14  & -1.55$\pm$0.14  & -1.43$\pm$0.17     & -1.57$\pm$0.13     & -1.54$\pm$0.13   & -1.57$\pm$0.13  & -1.48$\pm$0.19\\
           log([OIII]/H$\beta$)      & 0.77$\pm$0.01      & 0.77$\pm$0.01   & 0.75$\pm$0.01   & 0.78$\pm$0.01      & 0.77$\pm$0.01      & 0.77$\pm$0.01    & 0.73$\pm$0.01   & 0.73$\pm$0.02\\
           log([SII]/H$\alpha$)     & -1.01$\pm$0.01     & -1.03$\pm$0.01  & -0.92$\pm$0.02  & -1.02$\pm$0.05     & -1.00$\pm$0.01     & -1.01$\pm$0.01   & -0.88$\pm$0.03  & -1.07$\pm$0.12\\
\hline                        
\end{tabular}
\end{table*}

\begin{figure*}
   \centering
   \includegraphics[width=15cm]{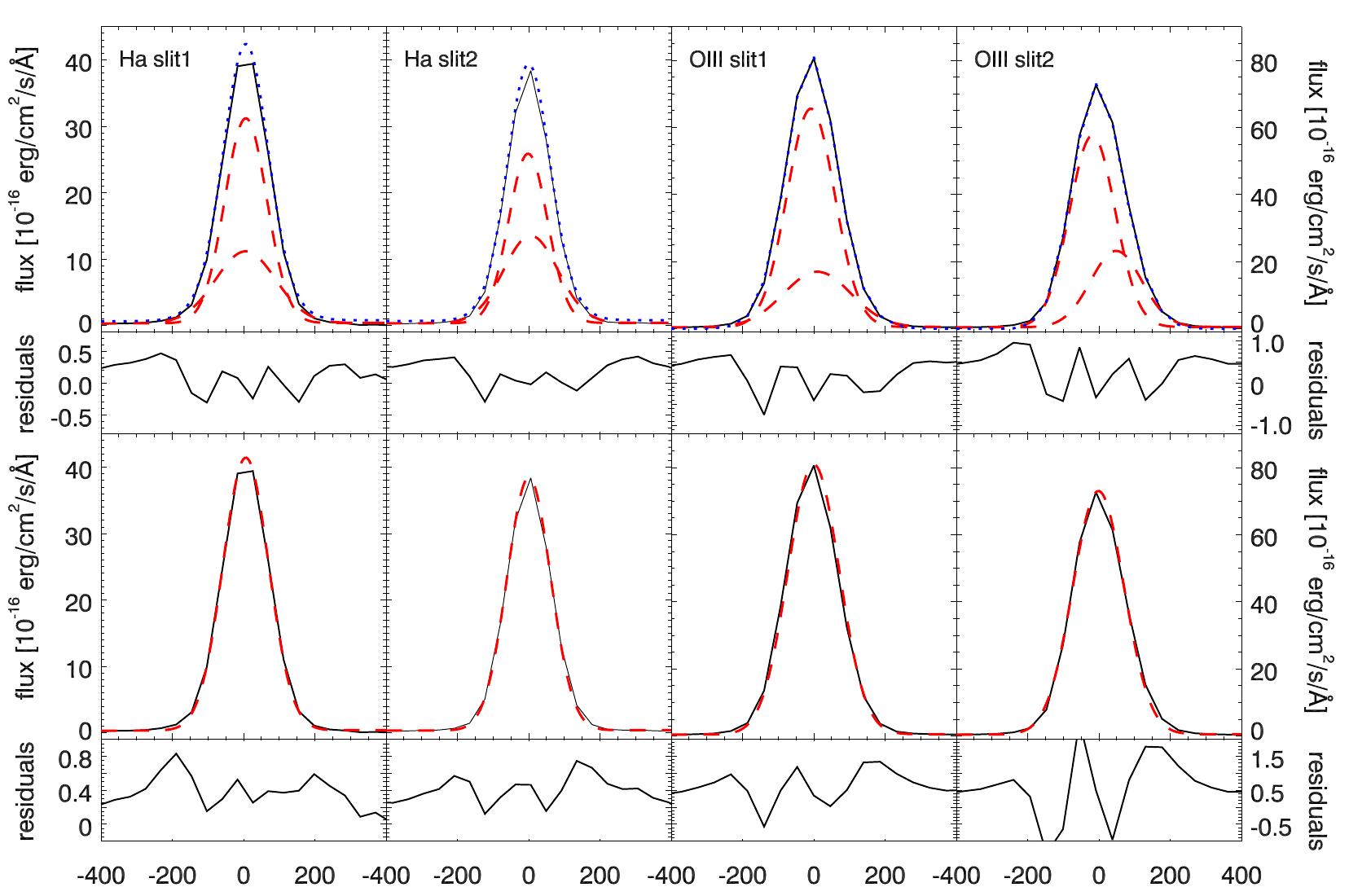}
         \caption{Top row: Fit to H$\alpha$ and [O\,III] in the integrated spectra of both slit positions with two Gaussians, a narrow and a broad component, and the residuals. Bottom row: The same lines but fitted with a single Gaussian. The residuals are larger than for the two-component fit and have a positive component in the wings, suggesting that a second, broad, component is actually present.}
         \label{fig:kinematics}
   \end{figure*}

\begin{figure*}
   \centering
   \includegraphics[width=\hsize]{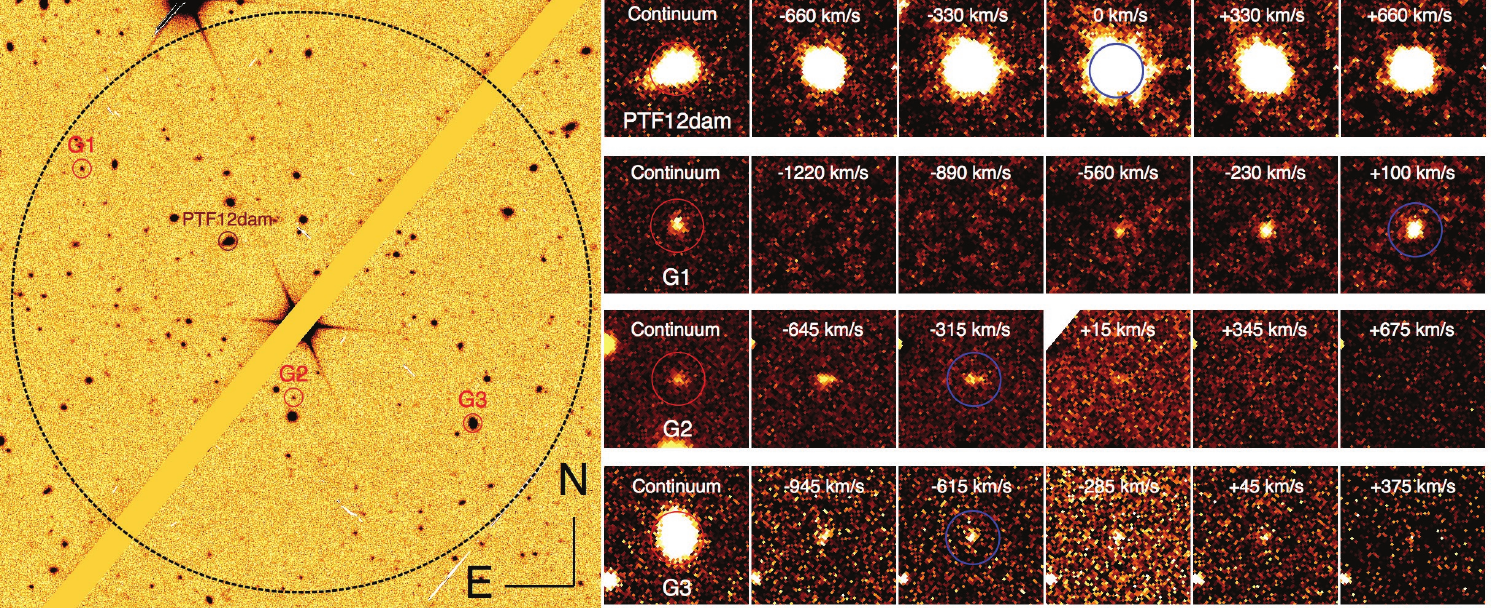}
         \caption{H$\alpha$ tunable narrow-band filter imaging of the field around PTF12dam. The galaxies with H$\alpha$ emission at redshifts similar to the host of PTF12dam are indicated in the finding chart taken with a continuum filter centered around 6800\AA{}. The sequences show the different galaxies in 5 steps of 8\AA{} around the center of H$\alpha$ in the host of PTF12dam. The distance in velocity is not the same for all galaxies due to the shift in wavelength across the FoV caused by the instrument.}
         \label{fig:TFs}
   \end{figure*}

\label{lastpage}


\begin{thebibliography}{}
\bibitem[Amor\'in et al.(2010)]{Amorin} Amor{\'{\i}}n, R.~O., P{\'e}rez-Montero, E., \& V{\'{\i}}lchez, J.~M.\ 2010, ApJ, 715, L128 
\bibitem[Amor\'in et al.(2012b)]{Amorin12} Amorin, R. et al. 2012, ApJ, 754, L22
\bibitem[Amor\'in et al.(2012a)]{AmorinGPabundances} Amor\'in, R., P\'erez-Montero, E., V\'ilchez, J. M \& Papaderos, P. \ 2012, ApJ, 749, 185
\bibitem[Amor\'in et al.(2014)]{AmorinEELGs} Amorin, R. et al. 2014, A\&A submitted, astro-ph/1403.3441
\bibitem[Brinchmann, Kunth \& Durret(2008)]{BKD08} Brinchmann, J., Kunth, D., \& Durret, F. 2008, A\&A, 485, 657,
\bibitem[Cepa et al.(2003)]{Cepa} Cepa, J., Aguiar-Gonzalez, M., Bland-Hawthorn, J., et al.\ 2003, proc SPIE, 4841, 1739 
\bibitem[Chen et al.(2014)]{ChenPTF12dam} Chen T.-W. et al. 2014, MNRAS submitted, astro-ph/1409.7728
\bibitem[Christensen et al.(2008)]{Christensen08} Christensen, L., Vreeswijk, P.~M., Sollerman, J., Th{\"o}ne, C.~C., Le Floc'h, E., \& Wiersema, K.\ 2008, A\&A, 490, 45
\bibitem[Cid Fernandes et al.(2005)]{CidFernandes05} Cid Fernandes, R., Mateus, A., Sodr{\'e}, L., Stasi{\'n}ska, G., \& Gomes, J.~M.\ 2005, MNRAS, 358, 363 
\bibitem[Chevalier et al.(2011)]{Chevalier11} Chevalier, R. A. \& Irwin, C. M. 2011, ApJ 729, L6
\bibitem[Firpo et al.(2011)]{Firpo11} Firpo, V., Bosch, G., H{\"a}gele, G.~F., D{\'{\i}}az, {\'A}.~I., \& Morrell, N.\ 2011, MNRAS, 414, 3288 
\bibitem[Gal-Yam et al.(2009)]{Gal-Yam09} Gal-Yam, A. et al. 2009, Nature, 462, 624
\bibitem[Gal-Yam(2012)]{Gal-YamSci} Gal-Yam, A. 2012, Science 337, 927
\bibitem[Gonz\'alez Delgado et al.(1999)]{GonzalezDelgado99} Gonzalez-Delgado, R.~M., Leitherer, C., 
\& Heckman, T.~M.\ 1999, ApJS, 125, 489 
\bibitem[Gonz{\'a}lez Delgado et al.(2005)]{GonzalezDelgado05} Gonz{\'a}lez Delgado, R.~M., Cervi{\~n}o, M., Martins, L.~P., Leitherer, C., \& Hauschildt, P.~H.\ 2005, MNRAS, 357, 945 
\bibitem[Gonz{\'a}lez Delgado et al.(2014)]{GonzalezDelgado14} Gonz{\'a}lez Delgado, R.~M. et al.\ 2014, A\&A, 562, A47 
\bibitem[Han et al.(2010)]{Han} Han, X.~H. et al.\ 2010, A\&A, 514, A24 
\bibitem[Izotov et al.(2007)]{Izotov07} Izotov Y. I., Thuan T. X., Guseva N. G., 2007, ApJ, 671, 1297
\bibitem[James et al.(2009)]{James} James, B.~L. et al.\ 2009, MNRAS, 398, 2 
\bibitem[Jaskot \& Oey(2013)]{JaskotOey} Jaskot, A. E. \& Oey, M. S., 2013, ApJ, 766, 91
\bibitem[Kasen\& Bildsten(2010)]{Kasen10} Kasen, D. \& Bildsten, L. 2010, ApJ, 717, 245
\bibitem[Kehrig et al.(2013)]{KehrigWR} Kehrig, C. et al.\ 2013, MNRAS, 432, 2731 
\bibitem[Kennicutt(1998)]{Kennicutt} Kennicutt, R. C. Jr. 1998, ARA\&A, 36, 189
\bibitem[Kewley et al.(2001)]{Kewley01} Kewley, L.~J., Heisler, C.~A., Dopita, M.~A., \& Lumsden, S.\ 2001, ApJS, 132, 37 
\bibitem[Kewley\& Ellison(2008)]{R23} Kewley, L.~J., \& Ellison, S.~L.\ 2008, ApJ, 681, 1183
\bibitem[Kuncarayakti et al.(2013)]{Kunca13} Kuncarayakti, H., Doi, M., Aldering, G., et al.\ 2013, AJ, 146, 31 
\bibitem[Leitherer et al.(1999)]{Leitherer99} Leitherer, C. et al.\ 1999, ApJS, 123, 3 
\bibitem[Levesque et al.(2010)]{Levesque10} Levesque, E.~M., Kewley, L.~J., Berger, E., \& Zahid, H.~J.\ 2010, AJ, 140, 1557 
\bibitem[Levesque et al.(2012)]{Levesque12} E. M. Levesque \& C. Leitherer 2013, ApJ, 779, 170 
\bibitem[Leloudas et al.(2014)]{LeloudasSUSHIES} Leloudas, G. et al., MNRAS, submitted, astro-ph/1409.8331
\bibitem[L\'opez-S\'anchez \& Esteban(2010)]{LopezWR} L\'opez-S\'anchez, A. R. \& Esteban, C. 2010, A\&A 516, id. A104
\bibitem[Lunnan et al.(2014)]{Lunnanhosts} Lunnan, R. et al. 2014, ApJ, 787, 138
\bibitem[Luridiana et al.(2014)]{Pynet} Luridiana, V., Morisset, C., \& Shaw, R.~A.\ 2014, arXiv:1410.6662 
\bibitem[Marino et al.(2013)]{Marino13} Marino, R.~A. et al.\ 2013, A\&A, 559, A114 
\bibitem[Mart\'in-Manj\'on et al.(2010)]{MartinManjon} Mart\'in-Manj\'on, M. L. et al. 2010, MNRAS, 403 2012
\bibitem[Meynet \& Maeder(2005)]{MeynetMaeder} Meynet, G. \& Maeder, A. 2005, A\&A, 429, 581
\bibitem[Nicholl et al.(2013)]{Nicholl13} Nicholl, M. et al. 2013, Nature, 502, 346
\bibitem[{\"O}stlin et al.(2008)]{Ostlin} {\"O}stlin, G., Zackrisson, E., Sollerman, J., Mattila, S., \& Hayes, M.\ 2008, MNRAS, 387, 1227 
\bibitem[Pastorello et al.(2010)]{Pastorello10} Pastorello, A., et al.\ 2010, ApJ, 724, L16
\bibitem[Pettini et al.(2004)]{Pettini04} Pettini, M., \& Pagel, B.~E.~J.\ 2004, MNRAS, 348, L59 
\bibitem[P{\'e}rez-Montero \& Contini(2009)]{PM09} P{\'e}rez-Montero, E., \& Contini, T.\ 2009, MNRAS, 398, 949 
\bibitem[P{\'e}rez-Montero et al.(2010)]{PerezMontero10} P{\'e}rez-Montero, E., Garc{\'{\i}}a-Benito, R., H{\"a}gele, G.~F., \& D{\'{\i}}az, {\'A}.~I.\ 2010, MNRAS, 404, 2037 
\bibitem[P{\'e}rez-Montero(2014)]{PM14} P{\'e}rez-Montero, E.\ 2014, MNRAS, 441, 2663 
\bibitem[Quimby et al.(2011)]{QuimbyNature} Quimby, R. M. et al. 2001, Nature 474, 487 
\bibitem[Quimby et al.(2012)]{QuimbyATEL} Quimby, R.~M. et al.\ 2012, The Astronomer's Telegram, 4121, 1 
\bibitem[Quimby et al.(2007)]{Quimby07} Quimby, R. M. et al. 2007, ApJ 668, L99
\bibitem[Sanchez Almeida et al.(2013)]{SanchezAlmeida} S\'anchez-Almeida, J. et al. 2013, ApJ, 767, 74
\bibitem[S{\'a}nchez Almeida et al.(2012)]{SanchezAlmeidagalaxies} S{\'a}nchez  Almeida, J., Terlevich, R., Terlevich, E., Cid Fernandes, R., \& Morales-Luis, A.~B.\ 2012, ApJ, 756, 163 
\bibitem[Sanders et al.(2012)]{Sanders} Sanders, N.~E. et al.\ 2012, ApJ, 758, 132 
\bibitem[Shirazi et al.(2012)]{Shirazi12} Shirazi, M., \& Brinchmann, J.\ 2012, MNRAS, 421, 1043 
\bibitem[Th\"one et al.(2011)]{ThoeneNature} Th\"one, C. C. et al. 2011, Nature 480, 72
\bibitem[Th{\"o}ne et al.(2008)]{Thoene08} Th{\"o}ne, C.~C. et al.\ 2008, ApJ, 676, 1151 
\bibitem[Th{\"o}ne et al.(2014)]{Thoene14} Th{\"o}ne, C.~C. et al.\ 2014, MNRAS, 441, 2034 
\bibitem[Vazdekis et al.(2010)]{Vazdekis10} Vazdekis, A. et al.\ 2010, MNRAS, 404, 1639
\end{thebibliography}
\end{document}